\documentclass[prl,superscriptaddress,twocolumn]{revtex4-1}

\usepackage{eurosym,amsmath,amssymb}
\usepackage{bm,graphicx}
\usepackage[usenames,dvipsnames]{xcolor}
\definecolor{gre}{rgb}{0.0, 0.4, 0.3}
\usepackage[colorlinks]{hyperref}
\usepackage[normalem]{ulem}

\hypersetup{colorlinks,citecolor=blue,linkcolor=blue,urlcolor=blue}

\begin{document}

\title{Chaotic cyclotron and Hall trajectories due to spin-orbit coupling}
\author{E.V. Kirichenko}
\affiliation{Institute of Physics, Opole University, Opole, 45-052, Poland}
\author{V. A. Stephanovich}
\affiliation{Institute of Physics, Opole University, Opole, 45-052, Poland}
\author{E. Ya. Sherman}
\affiliation{Department of Physical Chemistry, University of the Basque Country UPV/EHU,
48080 Bilbao, Spain}
\affiliation{IKERBASQUE Basque Foundation for Science, Bilbao, Spain}

\date{\today}

\begin{abstract}
We demonstrate that the synergistic effect of a gauge field, 
Rashba spin-orbit coupling (SOC), and Zeeman splitting can generate
chaotic cyclotron and Hall trajectories of particles. The physical origin of the chaotic behavior is that 
the SOC produces a spin-dependent (so-called anomalous) contribution to the particle velocity  
and the presence of Zeeman field reduces the number of integrals of motion. By using analytical  
and numerical arguments, we study the conditions of chaos emergence and report the dynamics both in the 
regular and chaotic regimes. {We observe the critical dependence of the dynamic patterns (such as the chaotic regime onset) on 
small variations in the initial conditions and problem parameters, that is the SOC and/or Zeeman constants. The transition to chaotic 
regime is further verified by the analysis of phase portraits as well as Lyapunov exponents spectrum.}  
The considered chaotic 
behavior can occur in solid state systems, weakly-relativistic plasmas, and cold atomic gases with synthetic 
gauge fields and spin-related couplings.
\end{abstract}

\maketitle

\section{Introduction}

Puzzling properties of chaotic motion in simple classical and quantum systems are among the most intriguing problems in modern physics. 
Recently observed features of a quantum chaos in cold gases \cite{frisch2014,Gao2015}, 
Rydberg excitons \cite{Assmann2016}, and polaritons \cite{Ostrovskaya2016} demonstrated that condensed matter is an excellent 
testbed for these studies ({see}, e.g., [\onlinecite{Schweiner2017}] and [\onlinecite{Schweiner2017a}]). They also posed 
new questions about relation between quantum chaos manifestations in  the spectrum and corresponding classical 
motion \cite{Gutzwiller,Reichl,Haake,Stockmann} in a broad variety of the systems. Therefore, it would be of 
interest to get insights into physical mechanisms underlying chaotic behavior. {This is especially true for the systems} with well-defined classical 
{(position and momentum) and quantum (e.g., spin)} degrees of freedom. Motivated by these results, we study 
two-dimensional (2D) motion of particles in magnetic fields with spin-orbit coupling (SOC) of the Rashba type \cite{Bychkov} 
and Zeeman splitting. These fields can be either genuine (electrons in solids or plasma \cite{Fabian,Dyakonov08}) or  
synthetic for cold atoms in designed coherent optical potentials
\cite{Dalibard,Zhaih2012,Spielman2013,Larson}. As an interesting example we mention that the spectra of
billiards \cite{Berggren,Csordas} and excitons \cite{Stephanovich} with SOC do not provide unambiguous relation to the 
classical chaos since their quantum chaotic features do not have classical counterpart, {although rectangular 
billiards with spin-orbit coupling driven by external electric fields clearly demonstrate a chaotic behavior \cite{Khomitsky2013}. 
Also, it is worth mentioning that a chaos existing in a host 2D system can have strong nontrivial effect 
on the spin transport there \cite{Liu2018}. Here we concentrate on the spin-orbit coupling effect on the emergence
of chaos in a simple semiclassical system. For this purpose we consider the diamagnetic effects in the orbital 
motion due to the Lorentz force and spin precession owing to the joint action of SOC and Zeeman effect. 
We have shown} that latter combination can lead to a chaotic behavior due to the anomalous spin-dependent 
contribution to the particle velocity \cite{Adams}. 
{This anomalous velocity in semiconductors is the core element of the phenomenon,
much resembling \textit{Zitterbewegung} (trembling oscillatory motion) of free relativistic electrons, 
described by Dirac equation. In semiconductor structures such as III-V quantum wells 
and wires, the \textit{Zitterbewegung} of electron wavepackets can be experimentally observed due to favorable 
energy and length scales \cite{Schliemann2005,Winkler2007}. We note also, that \textit{Zitterbewegung} - 
like motion plays an important role 
in cold atomic gases \cite{Zhaih2012}. }

The Hall effect, both in quantum and classical realizations, plays an important role in condensed matter physics. 
In a sufficiently strong magnetic field, the trajectory of a particle moving in a smooth 2D potential, 
resembles a closed narrow stripe in the vicinity of an equipotential line. This leads to the Hall effect 
quantization as the conductivity is solely due to the edge states. However, this simple picture does not 
take into account the SOC effects, which can strongly modify the motion and,  as a result, the entire 
cyclotron and Hall effect pictures. 

We demonstrate that at certain values of the Lorentz and electric forces and spin-dependent fields 
in terms of anomalous velocity and spin precession rate, the classical cyclotron and Hall trajectories 
become chaotic. With further increase in the Zeeman field, it becomes dominant and the spin dynamics 
turns regular. As a result, the effects of SOC decrease, and the chaos disappears although the particle 
trajectory can be strongly different from that without SOC. 
{As it is customary to chaotic systems, here we observe the critical dependence of the dynamics 
on both the initial conditions and problem parameters such as the SOC and Zeeman constants. Namely, we 
have shown, that system can enter and exit a stability domain by small variations of both initial 
conditions and above parameters.}

{To get further insights into emergent chaotic behavior, we analyze the phase portraits and 
so-called maximal Lyapunov exponent (MLE) $\lambda_{\max}$ both in deterministic and chaotic regimes. 
The MLE, being the largest value of the Lyapunov exponents spectrum, is often used as a marker of 
chaotic (if $\lambda_{\max}>0$) or a regular ($\lambda_{\max}<0$)  
behavior of dynamical system, see, e.g. \cite{gaspard}. Since we have shown that the chaotic (regular) 
trajectories obtained by numerical solution of systems of nonlinear differential equations 
correspond to positive (negative) MLE, the sign of $\lambda_{\max}$ is an additional consistency 
criterion of our numerical procedure.}

\section{Nonlinear equations of motion} 

We begin with the full Hamiltonian for a particle in an external electromagnetic field
characterized by time-independent vector-potential ${\mathbf A}=\left(A_{x},A_{y}\right)$:
\begin{equation}
H =\frac{p_{x}^{2}}{2}+\frac{p_{y}^{2}}{2}+
\alpha \left(p_{x}\hat{\sigma}_{y}-p_{y}\hat{\sigma}_{x}\right)
+H_{Z}+\varphi\left(\mathbf{r}\right), 
\label{Hfull}
\end{equation}
where $\mathbf{r}=(x,y),$ and 
\begin{equation} \label{pxy}
p_{x}\equiv -i{\partial}_{x}-A_{x},\qquad p_{y}\equiv -i{\partial }_{y}-A_{y},
\end{equation}
${\partial}_{x}\equiv\partial/\partial x,$ ${\partial}_{y}\equiv\partial/\partial y,$
$\alpha$ is the  SOC constant, 
$\hat{\sigma}_{i}$ are the Pauli matrices, $H_{Z}$ is the Zeeman term, and $\varphi(\mathbf{r})$ is the potential energy 
in the electric field.  {Hereafter we use the units with $\hbar=m=e=c=1$ and restore the physical units 
when discussing possible experimental implications of the results obtained.} 
Without loss of generality we take the magnetic field $B$  parallel to the $z-$axis:
\begin{equation}
\mathbf{B} ={\bm\nabla}\times {\mathbf A}=\partial_{x}A_{y}-\partial_{y}A_{x}. 
\end{equation}
The Zeeman contribution reads:
\begin{equation}
H_{Z}=\frac{\Delta _{x}}{2}\hat{\sigma}_{x}+\frac{\Delta_{z}}{2}\hat{\sigma}_z,
\end{equation}
where $\Delta_{x}$ and $\Delta_{z}$ are Zeeman splittings, which can, e.g., be produced by material 
magnetization, and will be assumed to be $B-$independent without loss of generality.

We derive the equations of motion for observables $O$ by using commutator-based approach \cite{land3}%
\begin{equation}
\dot{O}=i\left[ H,O\right].
\label{dOdt}
\end{equation}
Using commutation relation for operators $a$ and $b$ as $[a^{2},b]=a[a,b]+[a,b]a$ with $O=x$ and $y$,  
we obtain following expression for velocity in terms of time-dependent expectation values:
\begin{equation}\label{cz}
v_{x} =\dot{x}=p_{x}+\alpha \sigma _{y}; \qquad v_{y} =\dot{y}=p_{y}-\alpha \sigma _{x}. 
\end{equation}
The $\alpha \sigma_{y}$ and $-\alpha\sigma_{x}$ terms in Eq. \eqref{cz} correspond to so-called anomalous velocity, which is 
explicitly dependent on the spin components. This contribution appears due to SOC presence 
in the Hamiltonian \eqref{Hfull} and, as it will be demonstrated below, 
is responsible for the appearance of chaotic dynamics.

Then, applying Eq. \eqref{dOdt} for velocity in Eq. \eqref{cz} and spin components, 
we obtain the equations of motion:
\begin{equation} \label{lag}
\dot{v}_{x} =\omega_{c}v_{y}-\varphi_{x}(\mathbf{r})+\alpha \dot{\sigma}_{y}; \
\dot{v}_{y} =-\omega_{c}v_{x}-\varphi_{y}(\mathbf{r})-\alpha \dot{\sigma}_{x},
\end{equation}%
where {the cyclotron frequency in our units $\omega_c\equiv B$ and $\varphi_{x,y}(\mathbf{r})\equiv\partial_{x,y}\varphi(\mathbf{r})$. 
Equations \eqref{lag} should be augmented by those for spin evolution, caused by SOC and Zeeman terms in the form}
\begin{subequations} \label{pozh}
\begin{eqnarray}
\dot{\sigma}_{x} &=&2\alpha \left( v_{x}-\alpha \sigma _{y}\right)
\sigma _{z}-\Delta _{z}\sigma _{y}, 
\label{dsdt1}\\
\dot{\sigma}_{y} &=&\left( 2\alpha \left( v_{y}+\alpha \sigma
_{x}\right) -\Delta _{x}\right) \sigma _{z}+\Delta _{z}\sigma _{x}, 
\label{dsdt2}\\
\dot{\sigma}_{z} &=&-2\alpha \left( v_{x}\sigma _{x}+v_{y}\sigma
_{y}\right) +\Delta _{x}\sigma _{y}.
\label{dsdt}
\end{eqnarray}
\end{subequations}
The equations \eqref{lag} for accelerations and \eqref{pozh} for spin precession, 
being determined by the particle velocity, spin components, SOC, and magnetic field, are gauge-invariant 
since they do not include vector-potential explicitly. 

{Note that these equations are essentially semiclassical despite the quantum character of spin operators and 
similar to those of Ref. [\onlinecite{Larson}]. In the spirit of Ref. [\onlinecite{Larson}], 
they can be derived by using the Hamiltonian formalism of classical mechanics directly from \eqref{Hfull}. 
Namely, classical Hamiltonian equations for coordinate $r_{i}$ and momenta $p_{i}$ ($i=x,y$) components 
\begin{equation}\label{he}
\frac{dp_{i}}{dt}=-\frac{\partial H}{\partial r_{i}},\ \frac{dr_{i}}{dt}=\frac{\partial H}{\partial p_{i}},
\end{equation}
(where $H$ is Hamiltonian function \eqref{Hfull}) should be supplemented by those for expectation values 
of spin components (which we denote as $\sigma_{x,y,z}$ since they are essentially the same as those in 
Eq. \eqref{pozh}) obeying usual constraint 
\begin{equation} \label{spi}
\sigma_{x}^2+\sigma_{y}^2+\sigma_{z}^2=1.
\end{equation}
The constraint \eqref{spi} corresponds to the spin precession in the total
field given by the sum of the spin-orbit and Zeeman contributions. 
Latter equations yield exactly Eqs.\eqref{pozh} (with $v_{x,y}$ being substituted by sums defined in 
Eq. \eqref{cz}), while former ones are indeed Eqs. \eqref{lag}. It can be shown that the equations 
for spin components remain the same regardless of the derivation approach: either first commute the 
Hamiltonian \eqref{Hfull} with spin components according to the rule \eqref{dOdt} and then 
take expectation values or simply act within the classical approach \eqref{he} with respect to 
constraint \eqref{spi}.} 

{The equations \eqref{lag}, \eqref{pozh} } clearly demonstrate the unusual character of the system 
nonlinearity, consisting of two contributions. First one is constituted by the terms like $v_{a}\sigma_{b}$ ($a,b=x,y,z$) 
and second one is due 
to spin products $\sigma_{a}\sigma_{b}$. Both these contributions play an important role in the 
motion of the particle since the accelerations depend on the spin state, and, in turn, 
the spin evolution depends on velocity.
	
This geometrical constraint \eqref{spi}, additional to the energy 
conservation, makes the systems with SOC to be qualitatively different from
typical quantum and classical chaotic systems \cite{Gutzwiller,Reichl,Haake,Stockmann}.
In the absence of Zeeman and electric fields, the {time} evolution of  $z$ - components of the total angular momentum,
$L_{z}+\sigma_{z}/{2}$ {(where $L_{z}=xp_{y}-yp_{x}$)} is given by 
\begin{equation}
\frac{d}{dt}\left(L_{z}+\frac{\sigma_{z}}{2}\right)=-\frac{\omega_{c}}{2}\frac{d}{dt}r^{2}.  
\label{dLdt}
\end{equation}
This {SOC-independent} constraint with $L_{z}+{\sigma_{z}}/{2}=C-\omega_{c}r^{2}/2$,
where constant $C$ is determined by the initial conditions, strongly influences {the chaos emergence, making it less probable.} 

Although analytical investigation of {the above} system of nonlinear differential 
equations is not feasible, one can get a certain insight from a qualitative analysis as 
presented below. Namely, we trace possible chaotic behavior for two electric field realizations: 
zero field and uniform one $\varphi({\mathbf r})=Ey$, corresponding to cyclotron motion and Hall effect in the electric 
field ${\mathbf E}=-E{\mathbf y}$ ($\mathbf y$ is a unit vector in the $y$ direction), respectively.

\section{Chaotic cyclotron motion} 

For comparison with the conventional cyclotron motion, where
\begin{equation}\label{velocities}
v_{x}=v_{0}\sin(\omega_{c}t), \qquad v_{y}=v_{0}\cos(\omega_{c}t),
\end{equation}
we begin with solving the above equations \eqref{lag} and \eqref{pozh} at $\varphi({\mathbf r})\equiv\,0$
iteratively assuming the initial $\sigma_{z}(0)=1,$ $v_{x}(0)=0,$ and $v_{y}(0)=v_{0}.$
Qualitatively, the effect of SOC on the cyclotron motion is expected to be strong if 

\noindent (1) the typical anomalous velocity $\alpha$ [\onlinecite{Adams}]  is of the order of the initial velocity $v_{0}$ and 

\noindent (2) spin precession rate $2v_{0}\alpha$ is of the order of $\omega_{c}$ so that the trajectory radius 
should be of the order of spin precession length $1/\alpha$. 

Although for the Rashba coupling {without Zeeman field} 
the chaos does not appear due to the constraint (\ref{dLdt}), anisotropic  SOC [\onlinecite{Larson}] can lead to 
chaos as the latter constraint is lifted there. 

Since Zeeman field is essential in this case, we include it {in the form ${\bm\Delta}=(\Delta_{x},0,\Delta_{z})$} 
in our iterative procedure, presenting the velocity as ${\mathbf v}={\mathbf u}+{\mathbf V}$, where ${\mathbf u}$ 
is obtained in the "frozen spin" approximation with ${\bm\sigma}(t)=(0,0,1)$ and ${\mathbf V}$ 
is the corresponding correction. 
Substitution of the above iterative expression for the velocity into Eq. \eqref{lag} 
generates the following frozen-spin contribution {determined by the in-plane Zeeman field component $\Delta_{x}$:}
\begin{equation}
\left[ 
\begin{array}{c}
u_{x} \\ 
u_{y}%
\end{array}%
\right] =
\left(v_{0}-\frac{\alpha\Delta_{x}}{\widetilde{\omega}}\right)
\left[ 
\begin{array}{c}
 \sin {\widetilde{\omega}} t \\ 
\cos {\widetilde{\omega}} t%
\end{array}%
\right] +
\frac{\alpha \Delta _{x}}{\widetilde{\omega}}
\left[ 
\begin{array}{c}
0 \\ 
1
\end{array}%
\right], 
\end{equation}
where the renormalized frequency ${\widetilde{\omega}}=\omega_{c}+2\alpha^{2}$.
The equations for the ${\mathbf V}-$term {are determined by the out-of-plane $\Delta_{z}$ and read as:} 
\begin{subequations}
\begin{eqnarray}
\dot{V}_{x} &=&2v_{y}\alpha^{2}(\sigma _{z}-1)+2\alpha ^{3}\sigma_{x}\sigma _{z}+\alpha \Delta _{z}\sigma _{x}, \\
\dot{V}_{y} &=&-2v_{x}\alpha^{2}(\sigma _{z}-1)+2\alpha ^{3}\sigma_{y}\sigma_{z}+\alpha \Delta _{z}\sigma _{y}, 
\end{eqnarray}%
\end{subequations}
determine small-$t$ corrections due to Eqs.\eqref{pozh}:
\begin{subequations}\label{VxVy}
\begin{eqnarray}
V_{x} &=&v_{0}\alpha ^{2}\left[ 2\alpha ^{2}\omega_{c}+\Delta
_{z}-\left( \Delta _{x}-2\alpha v_{0}\right) ^{2}\right]\frac{t^{3}}{3}, \\
V_{y} &=&\alpha \left( 2\alpha ^{2}+\Delta _{z}\right) \left( 2\alpha
v_{0}-\Delta _{x}\right) \frac{t^{2}}{2}. 
\end{eqnarray}
\end{subequations}
{Equations \eqref{VxVy} demonstrate that to produce chaos, one needs Zeeman field component 
$\Delta_{x}$ of the order of $\alpha v_0$.}  

{Now, we can show that in strong Zeeman fields the chaos disappears and the motion returns to a regular behavior. 
As an example we take realization with $\Delta_{z}=0$ and $\Delta_{x}\gg\,\alpha v_{0}.$ For this realization the 
"spin part", i.e. Eqs. \eqref{pozh} acquire the form $\dot{\sigma}_{x}=0$, $\dot{\sigma}_{y}=-\Delta_{x}\sigma_{z},$ 
and $\dot{\sigma}_{z}=\Delta_{x}\sigma_{y}$ with the explicit solution 
\begin{equation}\label{ij1}
\sigma_{y}=-\sin(\Delta_{x}t),
\end{equation}
obtained with the above initial condition $\sigma_{z}(0)=1$, which implies  $\dot{\sigma}_{y}(0)=-\Delta _{x}$. 
Substitution of the solution \eqref{ij1} into the set \eqref{lag} generates following inhomogeneous system of equations for the 
velocity components
\begin{subequations}\label{ij2}
\begin{eqnarray}
\dot v_{x}&=&\omega_{c}v_{y}-\alpha\Delta_{x}\cos(\Delta_{x}t), \\
\dot v_{y}&=&-\omega_{c}v_{x}.
\end{eqnarray}
\end{subequations}
After solving it by the variation of constants with initial conditions $v_{x}(0)=0$, $v_{y}(0)=v_{0}$ , we finally arrive at:}
\begin{eqnarray}
\left[ 
\begin{array}{c}
v_{x} \\ 
v_{y}%
\end{array}%
\right] &=&
\left(v_{0}-\frac{\alpha \Delta_{x}\omega_{c}}{\omega_{c}^{2}-\Delta_{x}^{2}}\right) 
\left[ 
\begin{array}{c}
\sin \omega_{c} t \\ 
\cos \omega_{c} t%
\end{array}%
\right] \label{lappa} \\
&+&\frac{\alpha \Delta_{x}}{\omega_{c} ^{2}-\Delta_{x}^{2}}
\left[ 
\begin{array}{c}
\Delta_x\sin\left(\Delta_{x} t\right)  \\ 
\omega_{c}\cos\left(\Delta_{x} t\right) 
\end{array}%
\right], \nonumber
\end{eqnarray}
{The equation \eqref{lappa} defines double-periodic regular motion with, in general, a 
possible resonance between spin precession and cyclotron frequencies.}

\begin{figure}[h]
\begin{center}
\includegraphics*[width=0.9\columnwidth]{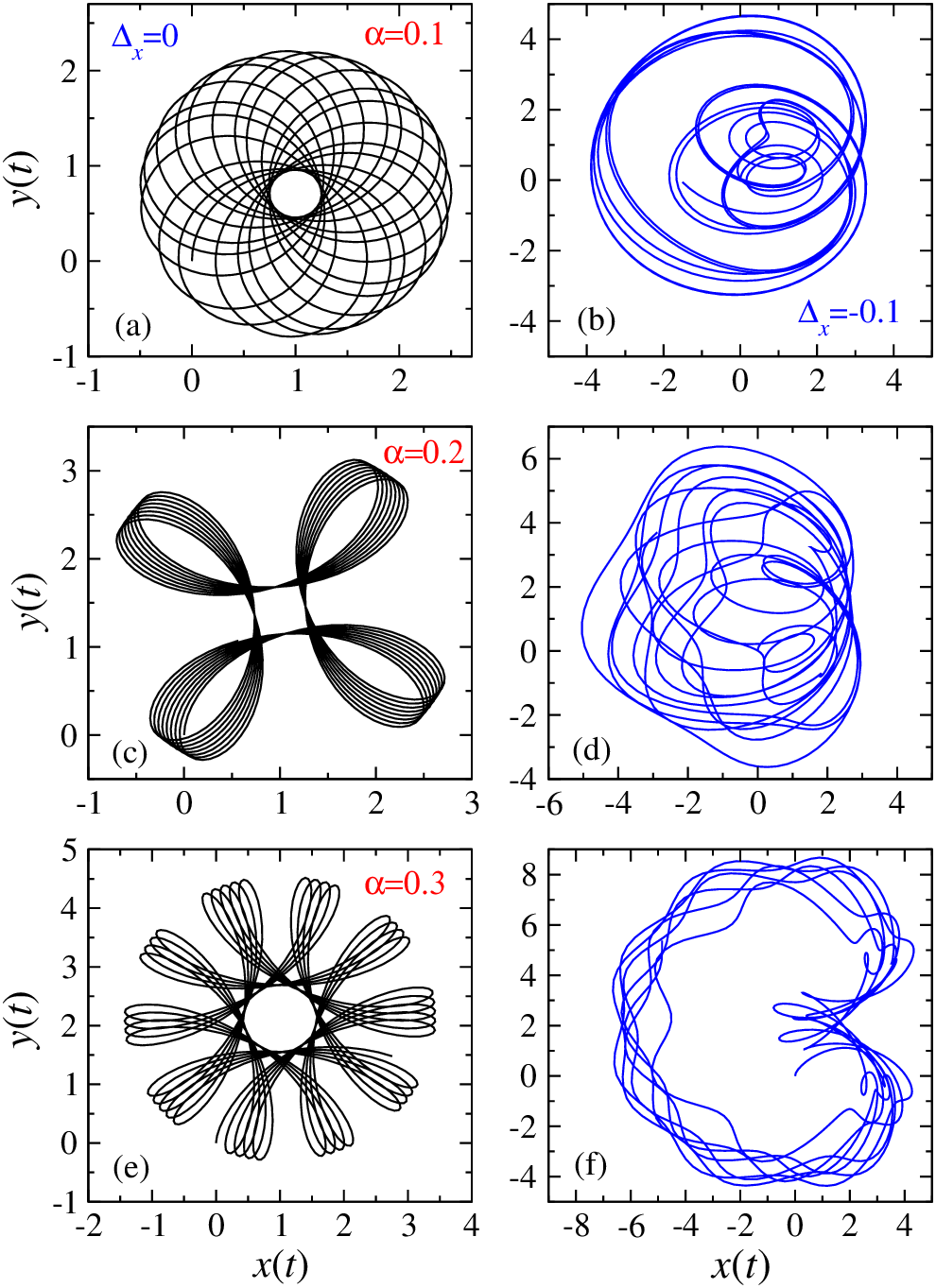}
\end{center}
\caption{ {Typical cyclotron trajectories for $t<10^{3}$ for the initial spin state $\sigma_{z}(0)=1$.
Left column (panels (a), (c), (e)) corresponds to Zeeman coupling  $\Delta_{x}=0$, right column (panels (b), (d), (f)) -
to $\Delta_{x}=-0.1$. Upper row (panels (a),(b)) corresponds to SOC constant $\alpha=0.1$, middle row (panels (c),(d)) corresponds to
$\alpha=0.2$ and lower row corresponds to $\alpha=0.3$. 
Here $\omega_{c}=0.1$, $\Delta_{z}=0,$ and the initial velocities $v_{x}(0)=0,v_{y}(0)=0.1.$ 
The absence of chaos at $\Delta_{x}=0$ (left column) is due to the constraint (\ref{dLdt}).}  } 
\label{fig:cyclotron}
\end{figure}

\begin{figure}[h]
\begin{center}
\includegraphics*[width=0.98\columnwidth]{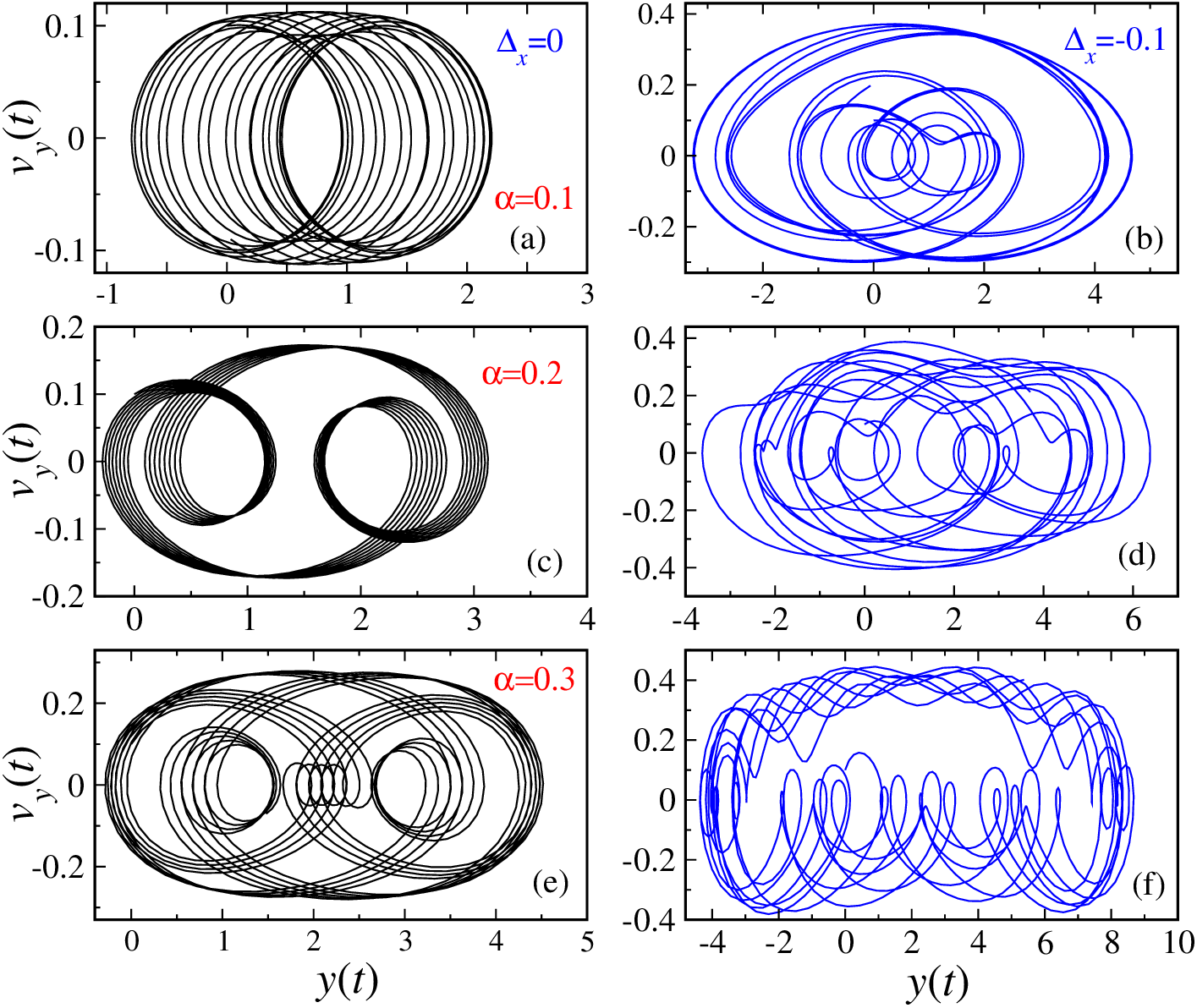}
\end{center}
\caption{{Phase portraits corresponding to the $(x,y)$ trajectories in Fig.\ref{fig:cyclotron} for the same set
of parameters and time intervals}.} 
\label{fig:cyclotronphase}
\end{figure}

\begin{figure}[h]
\begin{center}
\includegraphics*[width=0.9\columnwidth]{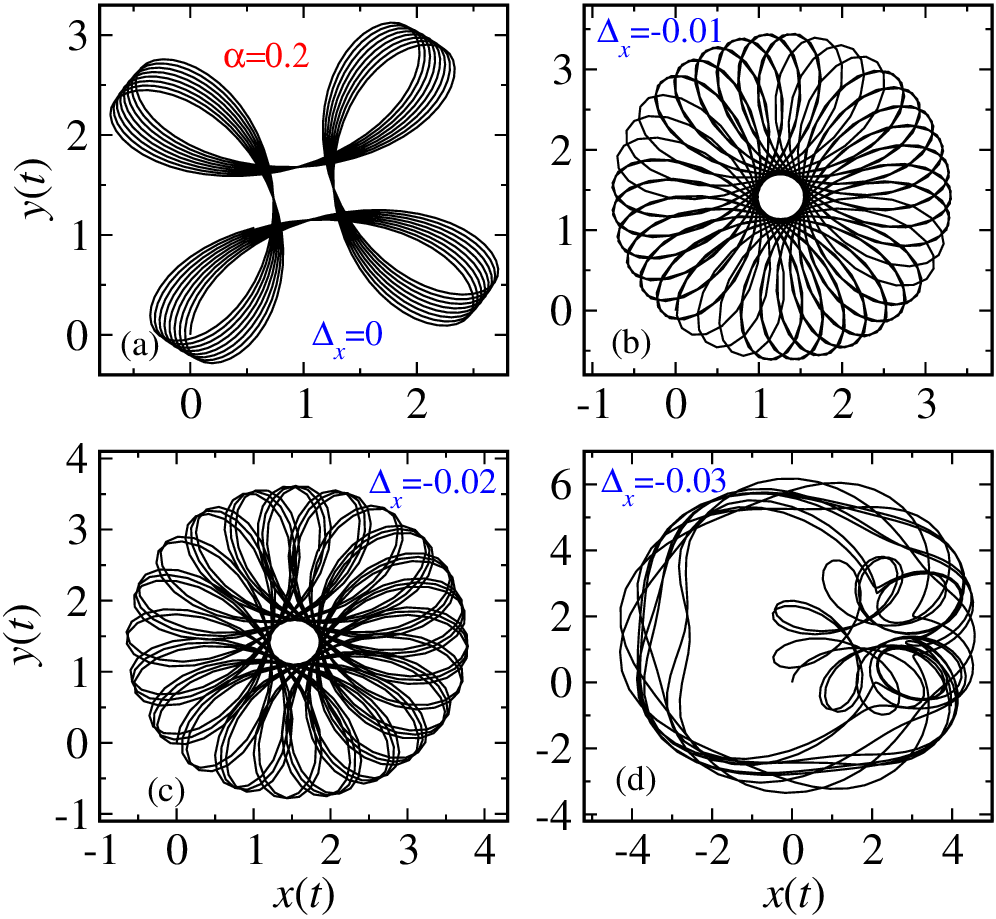}
\end{center}
\caption{{Strong dependence of the trajectories on the in-plane Zeeman field $\Delta_{x}$. The Figure illustrates transition 
from high density of the regular trajectories to their chaotization at small variation in the system parameters, 
as typical for chaotic systems.}}
\label{fig:deltadependence}
\end{figure}

\begin{figure}[h]
\begin{center}
\includegraphics*[width=0.98\columnwidth]{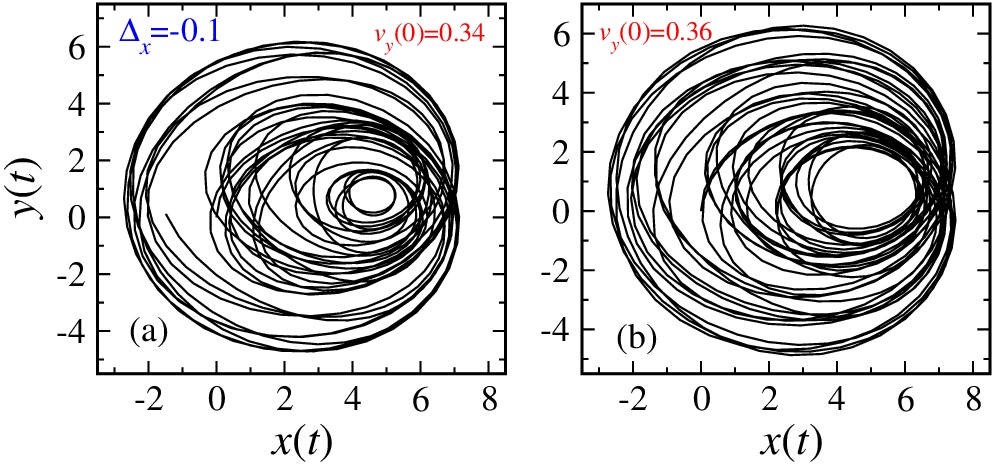}
\end{center}
\caption{{Transition from chaotic to regular high-density cyclotron trajectories at small
variation in the initial velocity $v_{y}(0)$. Parameters are the same as those in Fig. \ref{fig:cyclotron}(d)
($\Delta_{x}=-0.1$, $\Delta_z=0$, $\alpha=0.2$, $\omega_c=0.1$).}} 
\label{fig:vydependence}
\end{figure}

\begin{figure}[h]
\begin{center}
\includegraphics*[width=0.88\columnwidth]{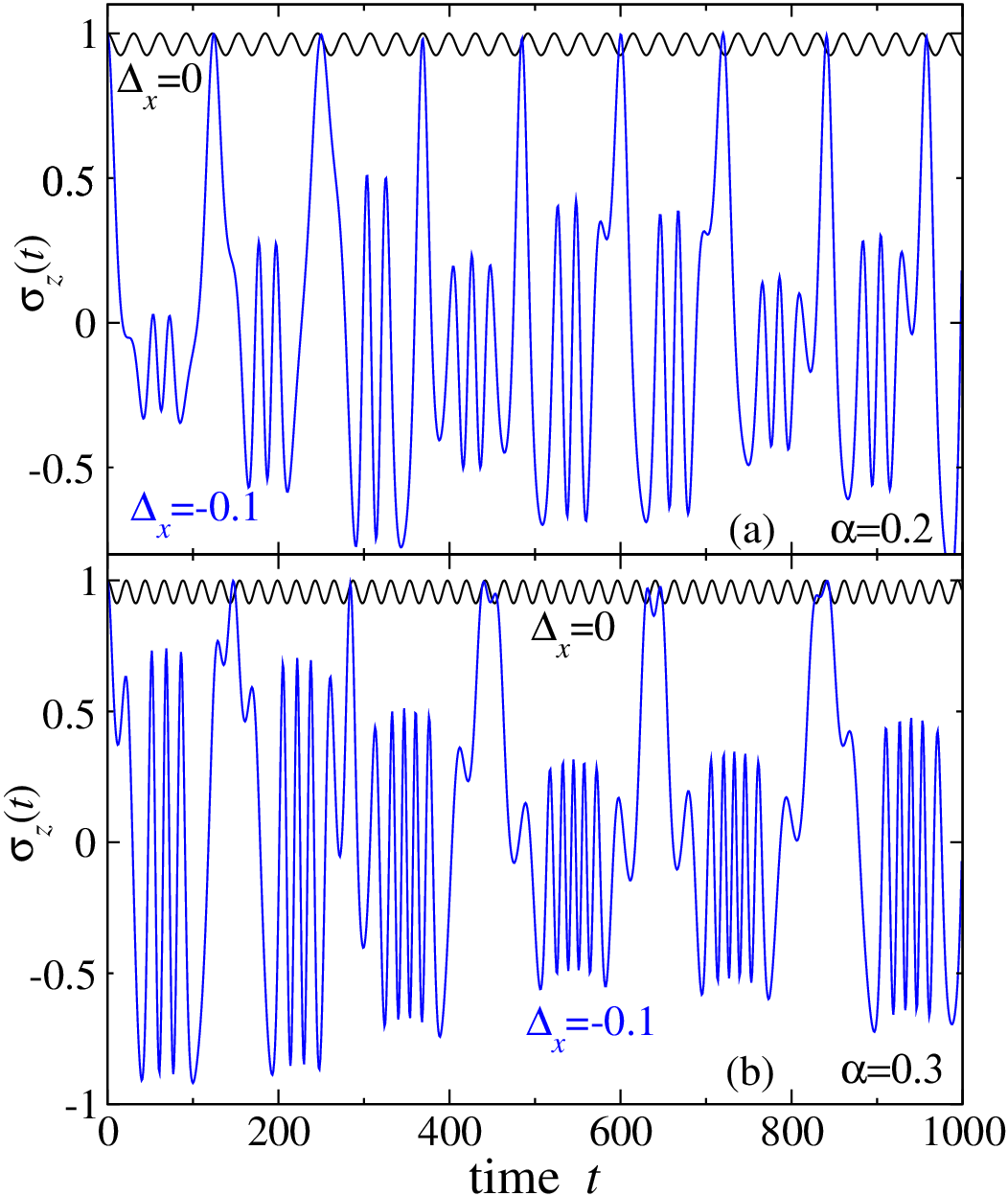}
\end{center}
\caption{Typical spin behavior for the cyclotron motion, shown in Fig. \ref{fig:cyclotron}. 
SOC constants are reported in the panels and the lines are marked with the values of $\Delta_{x}.$ 
Regular behavior of $\sigma_{z}(t)$ for $\Delta_{x}=0$ clearly corresponds to 
the quasiperiodic trajectories in Fig. \ref{fig:cyclotron}. Fast spin oscillations correspond to remote 
parts of the trajectories, while slow oscillations correspond to the nearby parts.}
\label{fig:cyclotronspin}
\end{figure}

{Since the full description 
of the system of interest requires the set $\left(\mathbf{r},\mathbf{v},{\bm\sigma},\dot{\bm\sigma}\right),$
with imposed constraints, the only way to depict it is to use projections of the above multidimensional 
surface onto specific planes as reported in Figs. \ref{fig:cyclotron} and \ref{fig:cyclotronphase}, presenting our main results
for the real space and phase trajectories.
It is seen that for regular real trajectories the phase ones are also regular, while in chaotic case, 
the phase portrait completely reflects the situation, being also chaotic.} 

{To characterize the chaotic trajectories in deterministic dynamic systems quantitatively, 
one usually introduces the Lyapunov spectrum (see, e.g., \cite{gaspard, maspard} and references therein),
providing a measure of the rate of time separation of initially (at $t=0$) infinitesimally 
close trajectories. For this purpose we first introduce vector 
${\mathbf Q}(t)=[\mathbf{r}(t),\mathbf{v}(t),{\bm\sigma}(t)]$. 
In this case, the trajectory separation $\delta {\mathbf Q}(t)$ characterizes how close are two trajectories 
at arbitrary time instant $t$, given that at $t=0$ they were almost the same, i.e. that the 
quantity $|\delta {\mathbf Q}_{0}|\equiv |\delta {\mathbf Q}(0)|\to 0$. Formally, at small separations, 
the entire set of our dynamical equations can be linearized to yield the sets of $ {\mathbf Q}_{0}^{[j]}$ 
and $\lambda_{j},$ with $|\delta {\mathbf Q}^{[j]}(t)|=e^{\lambda_{j} t}|\delta {\mathbf Q}_{0}^{[j]}|$, 
where $\lambda_{j}$ is the corresponding Lyapunov exponent. As the sets of initial conditions may be 
different (see Figs. \ref{fig:cyclotron}, \ref{fig:cyclotronphase}), the initial separation 
vectors $\delta {\mathbf Q}_{0}^{[j]}$ have different directions. This generates the entire spectrum of 
Lyapunov exponents, which in our case comprises 7 elements with $j=1,\ldots,7$. The most important 
characteristic of the spectrum is the Maximal Lyapunov exponent $\lambda_{\max}$ \cite{gaspard, maspard}, 
which determines if system is chaotic ($\lambda_{\max}>0$) or regular ($\lambda_{\max}<0$). 
One of the equivalent definitions of MLE, which we used in our calculations, reads \cite{maspard}
\begin{equation}\label{lapm}
\lambda_{\max}=\lim_{t \to \infty} \lim_{|\delta {\mathbf Q}_{0}| \to 0}\frac{1}{t}\ln\frac{|\delta {\mathbf Q}(t)|}{|\delta {\mathbf Q}_{0}|}.
\end{equation}
When the limit \eqref{lapm} is positive, the trajectories show extreme sensitivity to the initial conditions 
and the system becomes chaotic. Note that the limit $t \to \infty$ is taken in numerical procedure approximately and this makes 
the problem of $\lambda_{\max}$ calculation to take quite long time, especially in the chaotic regime. 
To calculate the Lyapunov spectrum for our problem, we used the algorithm of Ref. \cite{alg1} (see also Ref. \cite{alg2})
for implementation with Wolfram Mathematica software. 
Thus, we obtained, for Fig.  \ref{fig:cyclotron}(a) $\lambda_{\max}^{(a)}=-0.014$.  
Figures \ref{fig:cyclotron}(b) to (f) yield, respectively,  $\lambda_{\max}^{(b)}=0.027$, 
$\lambda_{\max}^{(c)}=-0.019$, $\lambda_{\max}^{(d)}=0.031$, $\lambda_{\max}^{(e)}=-0.025$, 
and $\lambda_{\max}^{(f)}=0.035$. 
This shows that the MLE's fully agree with the shape of the trajectories: 
they are positive for chaotic trajectories and negative otherwise. Hence, 
in our analysis, the MLE marker plays auxiliary role, confirming the appearance of the chaos 
for given sets of system parameters.}

{One of the main physical conclusions here follows from the comparison of the left and right columns of Figs. \ref{fig:cyclotron} 
and \ref{fig:cyclotronphase}. It is seen that solely SOC, even with a relatively large $\alpha$, 
does not generate chaos. To produce it, a Zeeman field is necessary. This is reported in the right columns of 
these Figures, where the chaotic trajectories are due to the interplay between the Zeeman and SOC fields.}

{To confirm the emergence of the chaos, we show other two peculiar features of the chaotic 
behavior such as strong dependence of the trajectories on the system parameters and initial conditions. 
Figure \ref{fig:deltadependence} shows the dependence on the in-plane magnetic field while Fig.\ref{fig:vydependence} 
demonstrates the dependence on the initial velocity. It is seen from Fig. \ref{fig:deltadependence}, that while at 
Zeeman splitting $\Delta_{x}=-0.02$, the system trajectory is still regular with $\lambda_{\max}=-0.012$, at 
a slightly smaller $\Delta_{x}=-0.03$ the system is already chaotic with $\lambda_{\max}=0.009$. Our calculations 
show that the same features occur also in other domains of $\Delta_{x}$ (for instance at $\Delta_{x}>0$) as 
well as of $\Delta_{y}$. Fig. \ref{fig:vydependence} reports the same instability with respect to $v_{y}(0)$: 
at a very small variation $0.34 < v_{y}(0) < 0.36$ the system passes from chaotic to regular behavior. This is confirmed by 
MLE calculations with $\lambda_{\max}=0.0032$ for $v_{y}(0)=0.34$ 
(Fig.\ref{fig:vydependence}(a)) and $\lambda_{\max}=-0.0087$ for $v_{y}(0)=0.36$ (Fig.\ref{fig:vydependence}(b)). 
It can be shown that the system is also sensitive to small variations in $v_{x}(0)$ as well as to all other possible 
combinations of initial conditions. }

To understand the spin evolution behind the regular and chaotic trajectories, we present in Fig. \ref{fig:cyclotronspin} 
the time dependence $\sigma_{z}(t)$ for four realizations of trajectories shown in Fig. \ref{fig:cyclotron}. As one can see
in the Figure, in the absence of the Zeeman coupling, spin shows relatively small deviations from its initial value,
corresponding to the above frozen spin approximation. The {spin behavior} in the absence of the Zeeman coupling is
consistent with the regular quasiperiodic trajectories in Fig. \ref{fig:cyclotron}. Indeed, for quasiperiodic 
trajectories the integral of velocity during one "period" is small. This smallness leads to a minute variation in 
the spin component $\sigma_{z}$ and, in turn, to regular trajectory, making the pattern consistent. At relatively large Zeeman splittings, the spin dynamics becomes chaotic, producing chaotic $\left(y(t),x(t)\right)$ trajectories. Nonzero Zeeman coupling $\Delta_{x}$ enhances the spin rotation, and, therefore, even if the particle displacement during one quasiperiod
is small, spin precession is essential for the orbital motion. In this case, the dynamics of $\sigma_z$ strongly modifies not only 
the effective cyclotron frequency $\omega_{c}+2\alpha^{2}\sigma_{z}$ but also the $\alpha$ - dependent terms in 
the equations of motion. The spin-orbit coupling here serves as a mediator between Zeeman-induced 
rotation and enhanced trajectory chaotization. On the contrary, $\Delta_{z}$ suppresses the spin rotation 
and stabilizes the trajectory against the chaos.
 
\section{Chaotic Hall effect} 

To compare the following results of approach with the conventional Hall effect in a uniform electric field $E\ll\,B$, 
we present the corresponding velocity as:
\begin{equation}
\left[ 
\begin{array}{c}
v_{x} \\ 
v_{y}%
\end{array}%
\right] =
u_{H}
\left[ 
\begin{array}{c}
-\cos \left(\omega_{c}t + \phi_{H} \right) \\ 
\hspace{0.4cm}\sin\left(\omega_{c}t + \phi_{H} \right)%
\end{array}%
\right] +v_{H}
\left[ 
\begin{array}{c}
1 \\ 
0
\end{array}%
\right],
\label{Hall}
\end{equation}
where $u_{H}=\left(v_{0}^{2}+v_{H}^{2}\right)^{1/2}$ and $\phi_{H}=\arctan(v_{0}/v_H)$. Here $v_{H}=-E/B$ is 
the conventional Hall velocity in the given geometry with $E>0.$ 
For the initial conditions $v_{x}(0)=0,v_{y}(0)=v_{0}$, 
chosen here {without loss of generality}, the Hall velocity in Eq. (\ref{Hall}) has the form
\begin{equation}
{v_{x} =v_{H}\left(1-\cos(\omega_{c}t)\right),} \quad v_{y} =v_{H}\sin (\omega_{c}t).
\label{vxvy}
\end{equation}
Note that here the effect of  SOC is stronger since the mean value of velocity 
during one cyclotron period is not small. Moreover, the constraint (\ref{dLdt}) is lifted here,
making the system prone to chaos even at $\Delta_{x}=0$. 

\begin{figure}[h]
\begin{center}
\includegraphics*[width=0.9\columnwidth]{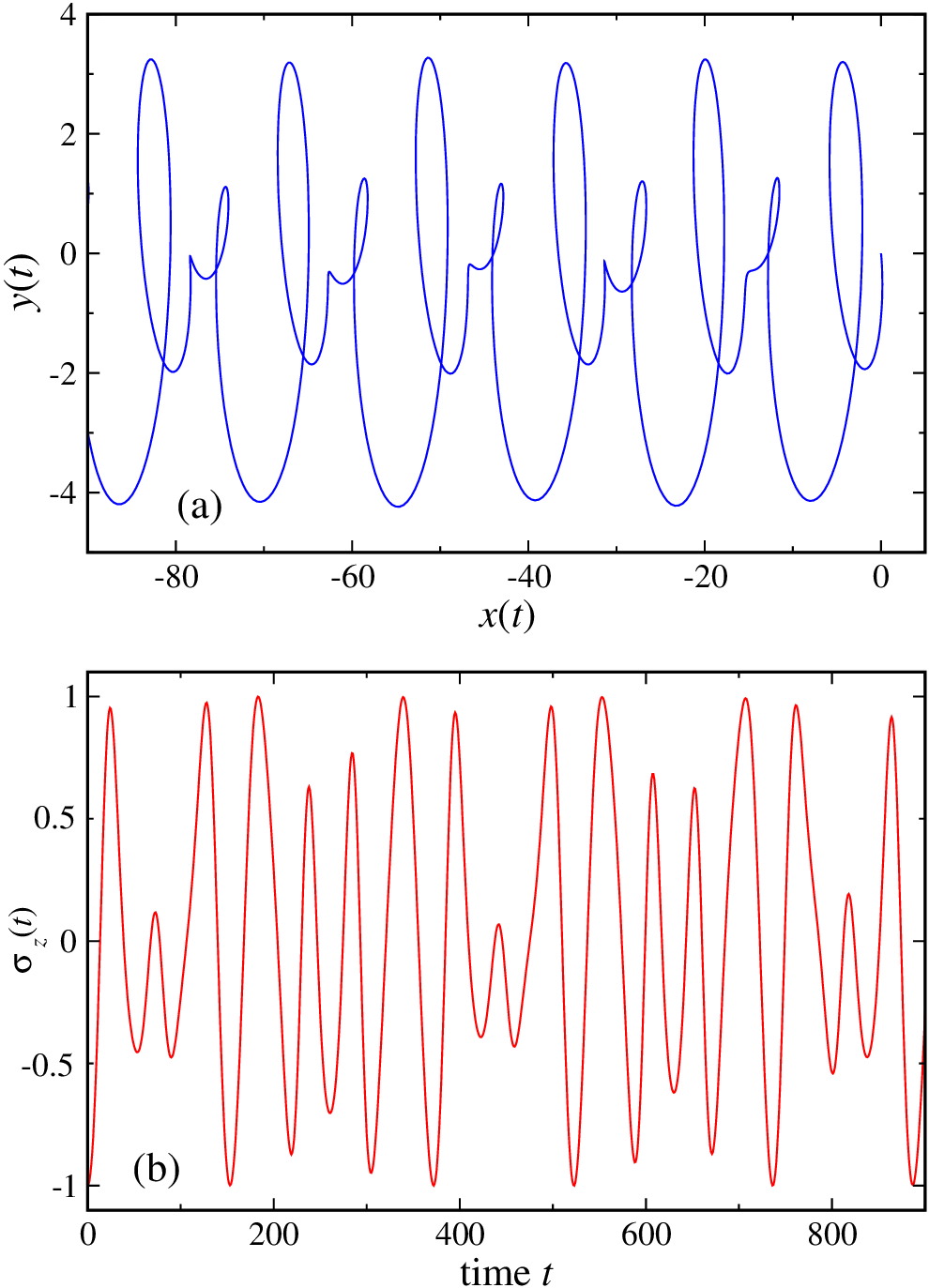}
\end{center}
\caption{ {Typical Hall trajectories (a) and spin evolution (b) for $\sigma_{z}(0)=-1$ and $\alpha=0.1$.
Here $\omega_{c}=0.1$, $\Delta_{z}=0$, $\Delta_{x}=0.1$, $E=0.01$ and the initial velocities $v_{x}(0)=v_{y}(0)=0$.} 
}
\label{fig:Hall1}
\end{figure}

\begin{figure}[h]
\begin{center}
\includegraphics*[width=0.9\columnwidth]{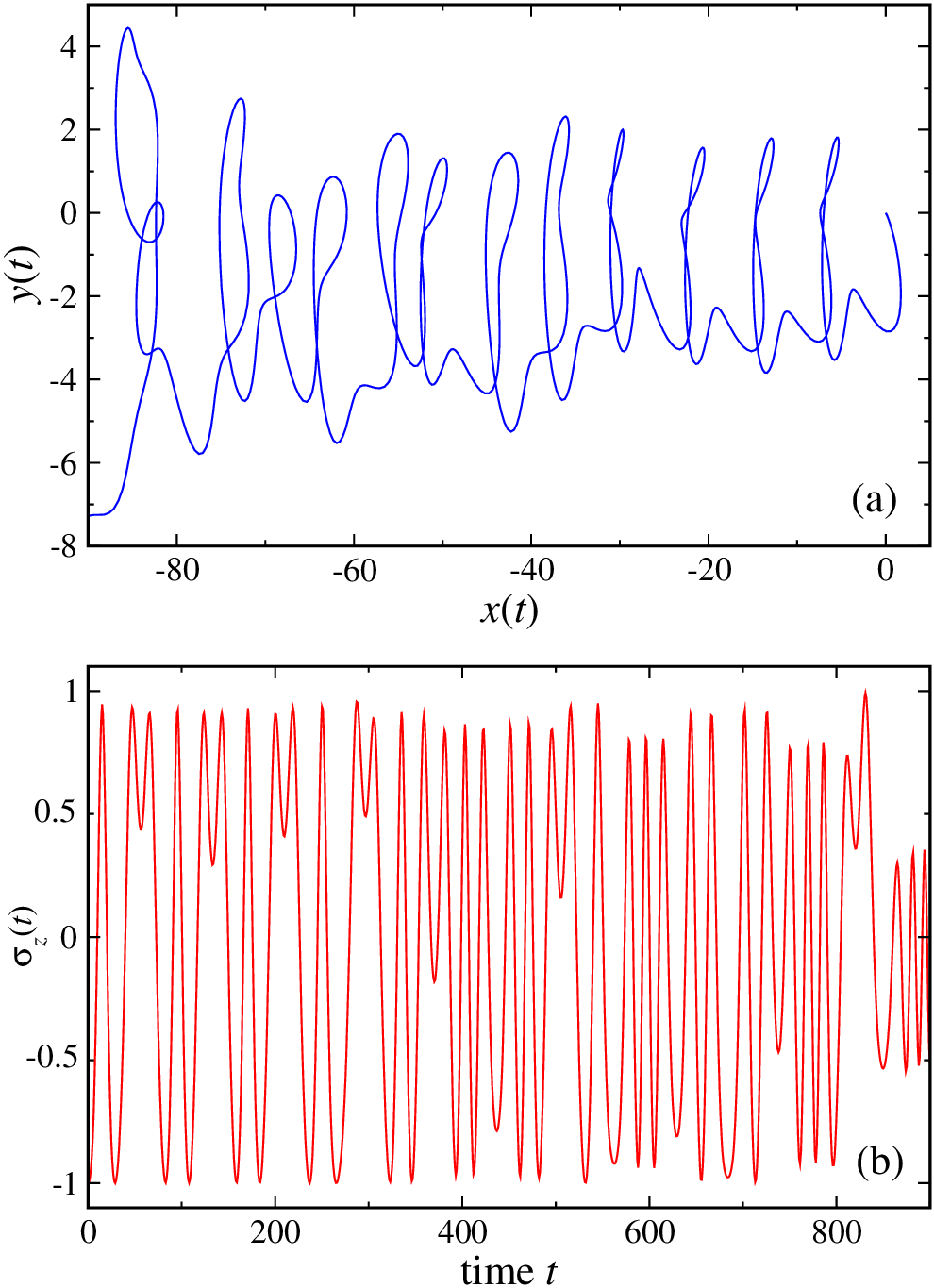}
\end{center}
\caption{ {Typical Hall trajectories (a) and spin evolution (b) for $\sigma_{z}(0)=-1$ and $\alpha=0.3$. 
Here $\omega_{c}=0.1$, $\Delta_{z}=0$, $\Delta_{x}=0.1$, $E=0.01$ and the initial velocities $v_{x}(0)=v_{y}(0)=0$.} 
}
\label{fig:Hall2}
\end{figure}

The numerically obtained trajectories and chaos development in the {Hall regime} are reported in Figs. \ref{fig:Hall1} and \ref{fig:Hall2}. 
It is seen, that at a relatively weak  SOC $\alpha=0.1$ (Fig. \ref{fig:Hall1}) the motion is approximately 
periodic being chaos-like within one period. At a stronger SOC $\alpha=0.3$  (Fig. \ref{fig:Hall2}), 
the periodicity disappears and the motion becomes truly chaotic. 

To analyze the qualitative effect of {the spin precession $\sigma_z(t)$}, we need to 
compare the renormalized by SOC cyclotron frequency $\omega_{c}+2\alpha^{2}\sigma_{z}$ to the electric field $E$. 
When $|\omega_{c}+2\alpha ^{2}\sigma _{z}|\gg E$, the system
is close to a conventional Hall effect. Otherwise, it is out 
of this regime, and the particle acceleration is determined {primarily} by the electric field. 
This occurs if $\alpha^{2}>\omega_{c}/2$ at time $t_{c}$ satisfying the condition  
$\sigma_{z}(t_{c})=-\omega_{c}/2\alpha^{2}$. 
The time $\tau$ the particle spends out of the {classical Hall regime} is 
of the order of $\tau\sim E/\alpha^{2}|\dot{\sigma}_{z}(t_{c})|$ if 
$\dot{\sigma}_{z}(t_{c})\ne 0$ or $\tau\sim \sqrt{E/|\ddot{\sigma}_{z}(t_{c})|}/\alpha$ 
if $\dot{\sigma}_{z}(t_{c})=0$. Accordingly, the velocity at this time interval 
has an increment $\delta v_{y}\sim -E\tau$. corresponding to 
elongation of trajectories along the $y$ - axis in Fig. \ref{fig:Hall2}. 
{Note that all above discussed regularities of chaotic behavior (such as sensitivity to initial 
conditions and/or problem parameters) take place for chaotic Hall effect as well.}

\section{Possible experimental implications} 

Now we are in a position to discuss system parameters required for observation {signatures of} the 
chaotic cyclotron motion and Hall effect for semiconductors and cold atoms. Note that the effects 
of  SOC on the regular cyclotron trajectories in semiconductors have been experimentally observed and 
theoretically studied in Ref. [\onlinecite{Rokhinson}]. The role of the anomalous spin-dependent velocity 
in the ac conductivity of 2D electron gas has been studied experimentally and theoretically 
in Ref. [\onlinecite{Tarasenko}]. {Full quantum mechanical analysis of the electronic  wave packets 
motion in magnetic field has been performed in Ref. [\onlinecite{Demikhovskii2008}]. 
While \textit{Zitterbewegung-}like effect has been clearly revealed and studied in details, 
no chaotic behavior appeared. The first reason is that the calculations have been made for the sets of parameters 
far away from the chaotic domains. The second reason is that the consideration in the paper \cite{Demikhovskii2008} 
is explicitly quantum mechanical with time-dependent expectation values being calculated with the help of 
corresponding wave functions. Although the relation between quantum \cite{Gutzwiller,Reichl,Haake,Stockmann} 
and classical chaos in spin-orbit coupling 
systems is very puzzling, such formalism, which does not deal with explicit time-dependent differential 
equations, would not, most probably, reveal features of the classical chaos. It is not excluded, however, that the approaches 
similar to \cite{Demikhovskii2008} may reveal some quantum chaotic features  
such as the energy levels repulsion, leading to non-Poissonian spectral statistics.}
  
Interesting features of the quantum Hall effect in the presence of {SOC} have been observed
experimentally in Ref. [\onlinecite{Shchepetilnikov2018}] and studied theoretically in Refs. 
[\onlinecite{Polyakov1995,Zarea2005,Pala2005,Hernangomez2014}].
It turns out also, that SOC term in the velocity is critically important for the spin Hall 
effect \cite{Sinova2015,Bi2013} and low-temperature transport \cite{Brosco}. 

Now we restore the physical units. To have a strong effect of spin-orbit coupling in the emerging chaotic behavior, 
we need to compare the cyclotron frequency $\omega_c=eB/mc$ with that corresponding to the shift of 
the conduction band bottom due to SOC, $\omega_{\rm so}=m\alpha^{2}/2\hbar^{3}$. 
As an example, we take the parameters for GaAs with $m=0.067m_{0}$ ($m_{0}$ is a free electron mass) and 
typical $\alpha = 10^{-6} \mbox{ meVcm}.$ This $\alpha$ corresponds to the 
anomalous velocity $\alpha/\hbar\approx 1.6\times\,10^{6}$ cm/s, and $2\omega_{\rm so}\approx 1.35\times 10^{11}\mbox{ s}^{-1}.$ 
Since for $B=0.1$ T, the corresponding $\omega_{c}\approx 2.6\times 10^{11}\mbox{ s}^{-1},$
we conclude that for chaos emergence, one needs either relatively weak magnetic fields 
or stronger SOC, which occur in In$_{x}$Ga$_{1-x}$As or InSb {2D} structures, albeit having smaller 
electron effective masses. Taking into account that at this field, $\omega_{c}$ and $\omega_{\rm so}$ are of 
the same order of magnitude, we also conclude that electron velocity $v_{0}\ge \alpha/\hbar$ and $\Delta_{x}/\hbar\sim 10^{11} \mbox{ s}^{-1}$ 
is sufficient to get strong effects of spin precession and chaos formation. In the Hall regime, the condition 
of fast precession has the form $\alpha m|v_{H}|/\hbar^{2}\omega_{c}\sim 1$, dependent on the electric 
field strengths. For the above values of $B$ and $\alpha$, this condition is satisfied at $v_{H}\sim 10^{7}$ cm/s. 
The situation is similar for cold atoms with synthetic SOC \cite{Zhaih2012,Larson}. Here the SOC energy, the typical kinetic energy, 
and the Zeeman term are of the same order of magnitude \cite{Spielman2013,Campbell}. Therefore, in the 
presence of a gauge field producing
a synthetic Lorentz force, the cold atoms motion is prone to chaos \cite{Larson}. 

\section{Conclusions} 

Two-dimensional materials and structures with spin–orbit coupling can exhibit a wealth of unexpected 
effects, both of fundamental physical interest and important for their possible electronic and spintronics 
applications \cite{Fabian}. In the present paper, using analytical and numerical arguments 
{in the semiclassical approximation}, we have demonstrated that joint effect of the Lorentz force, 
Zeeman splitting, and spin-orbit coupling {in 2D systems} generates chaotic 
trajectories of a particle moving in this combination of the fields.  
{ A typical chaotic trajectory can be described as a highly entangled path with high sensitivity to 
the small variations of initial conditions and/or system parameters. To describe this chaos mathematically, 
we utilize the phase portraits of the system under consideration as well as the spectrum of its 
Lyapunov exponents. The main role is played here by the MLE - the maximal exponent in the spectrum,
providing a consistency check for our numerical approach. Namely, for chaotic trajectories 
MLE is positive, while for regular ones it is negative. In our case, the reason for the chaos 
lies in the fact that the system loses integrability since it possesses only two integrals of 
motion for its phase space. Dynamically, this effect is clearly seen in the equations 
of motion including the anomalous spin-dependent velocity term.}

The specific physical mechanism behind the chaotization is the emergence of the spin-dependent term caused 
by the Rashba coupling in the effective Lorentz force related to the particle's velocity {and the \textit{Zitterbewegung}
effect.} In other words, the spin rotation in the Zeeman and Rashba fields is chaotically transformed into 
time-dependent anomalous (renormalized by spin degrees of freedom) velocity. {In this respect, 
our dependences, reported in Figs. \ref{fig:cyclotronspin} - \ref{fig:Hall2} can be considered 
as \textit{chaotic Zitterbewegung}. This interesting phenomenon needs further studies.} 
Therefore, the Zeeman field plays critical role since it can either trigger chaotization 
or suppress it, stabilizing the regular trajectories. As we have discussed in this paper, 
the considered effects are 
common for 2D semiconductor structures, weakly relativistic electrons and cold atoms with 
synthetic gauge, 
spin-orbit, and Zeeman couplings. The appearance of chaos in the Hall regime in smooth random potentials 
and dynamics of two-component wavepackets {in the domains of spin-orbit and Zeeman 
couplings suitable for the chaos emergence} are of interest and will be studied separately. 
{In addition, generalization of the proposed approach for the spin-orbit coupled Bose-Einstein
condensates \cite{Wang,Fujimoto} and cold atomic gases \cite{Lan} with the effective (pseudo)spin $s=1,$ 
demonstrating a more classical behavior than $s=1/2,$
can reveal possibly chaos-related properties of these systems.}

\begin{acknowledgments}

E.K. and V.S. acknowledge support of the Narodowe Centrum Nauki in Poland as research Project No. DEC-2017/27/B/ST3/02881.
E.S. acknowledges support of the Spanish Ministry of Science and the European Regional Development
Fund through PGC2018-101355-B-I00 (MCIU/AEI/FEDER,UE), and the Basque Government through
Grant No. IT986-16.

\end{acknowledgments}

\end{document}